\documentclass[runningheads]{llncs}
\usepackage{graphicx}
\usepackage{subcaption}
\usepackage{graphicx}
\usepackage{algorithm}
\usepackage[noend]{algpseudocode}
\usepackage{listings}
\usepackage{soul}
\usepackage{xcolor}
\usepackage{amsmath}
\usepackage{subcaption}
\usepackage{multirow}
\usepackage{makecell}
\usepackage{booktabs}
\usepackage{svg} % svg images
\usepackage{caption}
\usepackage{siunitx}
\usepackage{indentfirst} % auto indentation
\usepackage{comment}
\usepackage{amsmath} % to align equations
\usepackage{multirow} % to merge table cells
\usepackage{tipa}
\usepackage{tipx} % international phonetic alphabet characters
\usepackage{siunitx} % percentages
\usepackage{ctable}

\apptocmd{\thebibliography}{\scriptsize}{}{}

\definecolor{blue}{rgb}{0,0,1}

\definecolor{mygreen}{rgb}{0,0.6,0}
\definecolor{mygray}{rgb}{0.5,0.5,0.5}
\definecolor{mymauve}{rgb}{0.58,0,0.82}
\definecolor{mygreen2}{rgb}{0,0.4,0}
\lstdefinestyle{myCustomCppStyle}{
    language=C++,
    stepnumber=1,
    numbersep=10pt,
    tabsize=4,
    showspaces=false,
    showstringspaces=false,
    keywordstyle=\color{mymauve},
    commentstyle=\color{mygreen2},
    stringstyle=\color{mygreen},
    basicstyle=\fontsize{6.5}{!}\ttfamily,
    breaklines=true,
    frame=single,
    postbreak=\mbox{\textcolor{mygreen2}{$\hookrightarrow$}\space},
    morecomment=[l][\color{mygray}]{\#}
}
 
\begin{document}
\title{A Study of Performance Portability in Plasma
Physics Simulations}%\thanks{Supported by organization x.}}
\author{Josef Ruzicka\inst{1,2}\orcidID{0009-0003-4423-2612} \and
Christian Asch\inst{1}\orcidID{0000-0002-3111-4858} \and
Esteban Meneses\inst{1,2}\orcidID{0000-0002-4307-6000} \and Markus Rampp\inst{3}\orcidID{0000-0001-8177-8698} \and Erwin Laure\inst{3}\orcidID{0000-0002-9901-9857}}
\authorrunning{J. Ruzicka \textit{et al.}}
\institute{National Advanced Computing Laboratory, National High Technology Center, San José, Costa Rica\\
\email{{\{jruzicka, casch, emeneses\}@cenat.ac.cr}}\\ \and  School of Computing, Costa Rica Institute of Technology, Costa Rica \\ \and
Max Planck Computing and Data Facility, Max Planck Society, Garching, Germany\\
\email{\{markus.rampp, erwin.laure\}@mpcdf.mpg.de}}
\maketitle              % typeset the header of the contribution
\begin{abstract}
    The high-performance computing (HPC) community has recently seen a substantial diversification of hardware platforms and their associated programming models. From traditional multicore processors to highly specialized accelerators, vendors and tool developers back up the relentless progress of those architectures. In the context of scientific programming, it is fundamental to consider performance portability frameworks, \textit{i.e.}, software tools that allow programmers to write code once and run it on different computer architectures without sacrificing performance. 
We report here on the benefits and challenges of performance portability using a field-line tracing simulation and a particle-in-cell code, two relevant applications in computational plasma physics with applications to magnetically-confined nuclear-fusion energy research. For these applications we report performance results obtained on four HPC platforms with server-class CPUs from Intel (Xeon) and AMD (EPYC), and high-end GPUs from Nvidia and AMD, including the latest Nvidia H100 GPU and the novel AMD Instinct MI300A APU. Our results show that both Kokkos and OpenMP are powerful tools to achieve performance portability and decent ``out-of-the-box" performance, even for the very latest hardware platforms. For our applications, Kokkos provided performance portability to the broadest range of hardware architectures from different vendors.   

\keywords{Parallel Programming \and Performance Portability \and Plasma Physics.}
\end{abstract}
%
%
%

% introduction
\vspace{-2.5em}
\section{Introduction}
\vspace{-.5em}
As the energetic and environmental needs of humanity call for a turning away from fossil fuels, clean and renewable energy sources with little effect on global warming, such as nuclear fusion, receive increasing attention in research and society~\cite{Freidberg2007-lm}.
Fusion energy research, in particular, is well known to require vast amounts of computational power for modeling the complex plasma-physics phenomena spanning large ranges of spatial and temporal scales.
Due to the long-term efforts invested in developing the simulation codes, it is highly desirable to write them in a performance-portable way, \textit{i.e.}, allowing the reuse of the same source code on different high-performance computing (HPC) systems and without sacrificing application performance.

Commonly, the ``MPI+X" programming paradigm is adopted, which employs the Message Passing Standard (MPI) for handling parallelism across nodes, complemented by some other model (X) for intra-node parallelism (including accelerators), which is typically relying on some sort of shared memory.  
One of the most relevant models in this respect is OpenMP, a directive-based approach to parallelism that also supports offloading computation to accelerators while maintaining compatibility for shared memory parallelism in the CPU~\cite{gayatriCaseStudyPerformance2019}, enabling portability across computing architectures. It is a mature standard that requires a compiler capable of interpreting the directives and also provides additional functionality through a library API, allowing deeper optimizations. % \MJR{what does this mean: used as a run-time library in the source code}
Another framework that focuses on performance portability is Kokkos~\cite{9485033}, which is implemented as a library and enables portability of C++ codes between CPUs and GPUs. It allows to use the same data structures for different devices (\textit{e.g.} CPU or GPU) by providing a polymorphic data layout that adjusts to the device that is being used, thereby, for example, addressing the often-encountered design decision of array of structures (AoS) versus structure of arrays (SoA), none of which leading to a memory-access pattern that is well-suited for different kinds of device at the same time. These polymorphic data structures are called Views and are described as ``a potentially reference counted multidimensional array with compile-time layouts and memory space". In other words, they account for blocked or coalesced data access patterns based on the specified memory and execution spaces defined for multicore or GPU systems, respectively~\cite{KokkosAPI,CarterEdwards20143202}.

%"Kokkos is unique in that (1) it is purely a library approach, and (2) it enables portability to CPUs and GPUs, and (3) it provides polymorphic data layout. 
%On a CPU a computational kernel should have blocked data access pattern; however, on a GPU the computational kernel should have a coalesced data access pattern. This conflicting data access pattern requirement is commonly referred to as the array of structures (AoS) versus structure of arrays (SoA) problem."
%https://www.sciencedirect.com/science/article/abs/pii/S0743731514001257

% We document the experience of integrating Kokkos into the software and provide advice that can be helpful when doing similar work
In this paper, we evaluate the Kokkos framework using two different plasma-physics simulation codes, a particle-in-cell (PIC) code and a field-line tracer code. We compare the computational performance of the Kokkos code variants with the corresponding baseline implementations on four HPC systems using two different CPU models and five different GPU models.

% TODO JOSEF

%, and finally we test the simulations on GPU with the same Kokkos code with minimal modifications.

Previously, our team developed BS-SOLCTRA~\cite{Jimenez2019BS-SOLCTRA}, an OpenMP code for multicore systems for tracing field-lines that is used for simulating different stellarator coil configurations. The data structures for representing the particles were based on the AoS approach. This work was later continued by porting the code to GPUs, still using OpenMP. However, for performance reasons, this required transforming the data layout to one-dimensional arrays, thereby compromising portability (in the sense of a single source code) between architectures, as the new one-dimensional array variant for the GPU showed the expected underperformance on the CPU. This experience specifically motivated us to consider Kokkos with its polymorphic data structures and its promise of maintaining a single-source code by transparently accounting for the most appropriate memory-access pattern on both CPUs and GPUs~\cite{10.1007/978-3-031-23821-5_3}. Our assessment is complemented and broadened by considering, in addition, a simulation code based
on the particle-in-cell (PIC) method, which is one of the most commonly used methods in computational plasma physics.
%\MJR{what about the motivation for PIC ?}
%For our Kokkos implementations we make use of Unified Virtual Memory because of the flexibility it brings when accessing memory from different processing units. In addition, we have conducted tests on the model in various HPC plasma physics simulations related scenarios~\cite{10.1007/978-3-031-23821-5_3}.

%These versions have been tested on different clusters and different computer architectures
The contributions of this paper are two-fold. First, we develop performance-portable variants of our field-line tracer code, BS-SOLCTRA, and a representative PIC code. Second, we offer a comparative evaluation of Kokkos and OpenMP for the different variants of the code.

%\MJR{can we qualify: is this a specific PIC code, or a representative one, or a relevant one, \dots?}. \MJR{this sentence is not clear: Second, we offer a comparative evaluation of two leading performance-portable tools and use them to implement our plasma physics code.}

%The remainder of this paper is structured as follows: Section II reviews some related work on performance portability and plasma physics simulations. In Section III, we describe our base implementations and the way we adapt them to be performance portable with Kokkos. Section IV provides details on the experimental setup and design, as well as the results obtained when working on a single node and scaling on multiple nodes. Finally, Section V issues the insights and key findings of our work.
The remainder of this paper is structured as follows: Section 2 describes the numerical methods used in our plasma physics simulations, provides the definitions used for performance portability, and lists some related work on performance portability and plasma physics simulations. In Section 3, we describe our baseline implementations of the two codes and the way we adapt them to Kokkos. Section 4 provides details on the setup, and the results of our performance-evaluation. Finally, Section 5 summarizes insights and key findings of our work.

% we test the simulations on diverse architectures with the same Kokkos code with minimal modifications.
% We document the experience of porting the software to Kokkos, testing it on different architectures with minimal code modifications,...

% Collaboration between scientists from different laboratories commonly mandates the simulations to be executed on different computers with mismatching architecture, in consequence, having performance portable simulations, that is to say, simulations that can be run in different platforms without losing significant performance, becomes very valuable as it allows the scientists to analyze and share results in a time that would otherwise be spent on code refactoring.

% Research questions:\\
% How well the performance portable implementation scales compared to the base version?\\

% How portable is a Kokkos implementation of plasma simulations across architectures and across supercomputing platforms?\\
% How expressive is Kokkos when porting pre-existing applications? What are the limitations/advantages of Kokkos?\\
% What are the benefits of performance portability for research collaboration?

% background
\vspace{-1.em}
\section{Background}
%\vspace{-0.35em}

%\subsection{Plasma Physics Simulations}
\vspace{-.5em}
\subsection{Basics of the Numerical Models}
%Researchers rely on computational simulations designed to study plasma kinetic dynamism using numerical methods that approximate mathematical equations that would otherwise be tremendously complicated to solve.
%Researchers rely on computational simulations that use numerical methods to approximate mathematical equations in order to study plasma kinetic dynamism.

%\begin{figure}
%    \centering
%    \includegraphics[width=1\textwidth]{figures/Sim-Diagrams.jpeg}
%    \captionsetup{justification=centering}
%    \caption{From left to right, Field line tracer simulation diagram, PIC simulation diagram. Both as 2D representations}
%    \label{fig:simulation diagrams}
%\end{figure}
\vspace{-1.5em}
\begin{figure}[]
  \centering
  \begin{subfigure}[b]{0.25\textwidth}
    \includegraphics[width=\textwidth]{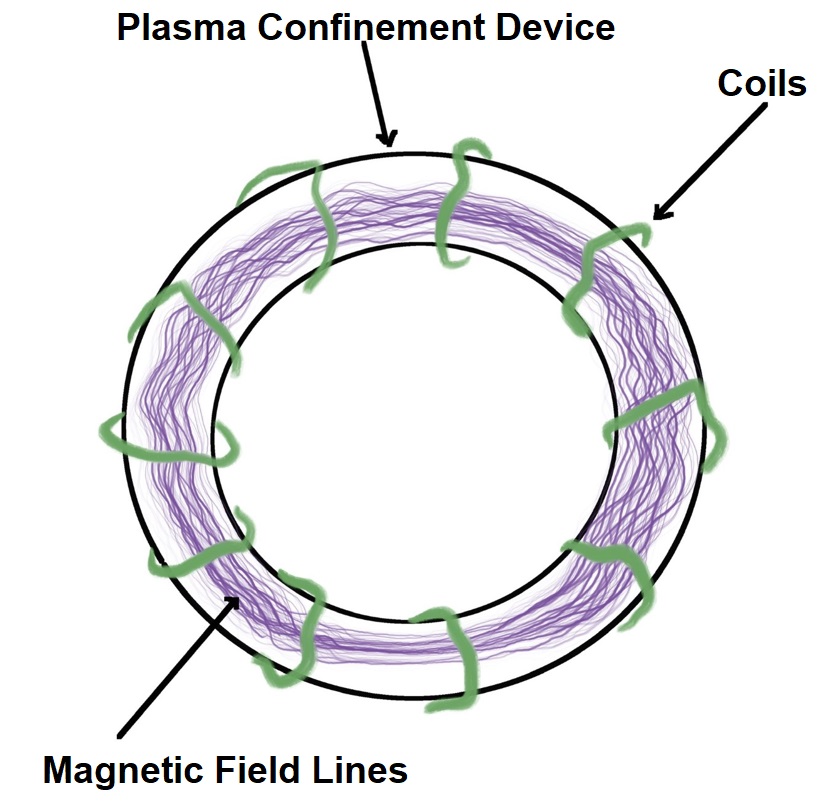}
    \caption{Field-line tracer}
    \label{fig:stellarator}
  \end{subfigure}
  %\hfill
  \hspace{3em}%
  \begin{subfigure}[b]{0.25\textwidth}
    \includegraphics[width=\textwidth]{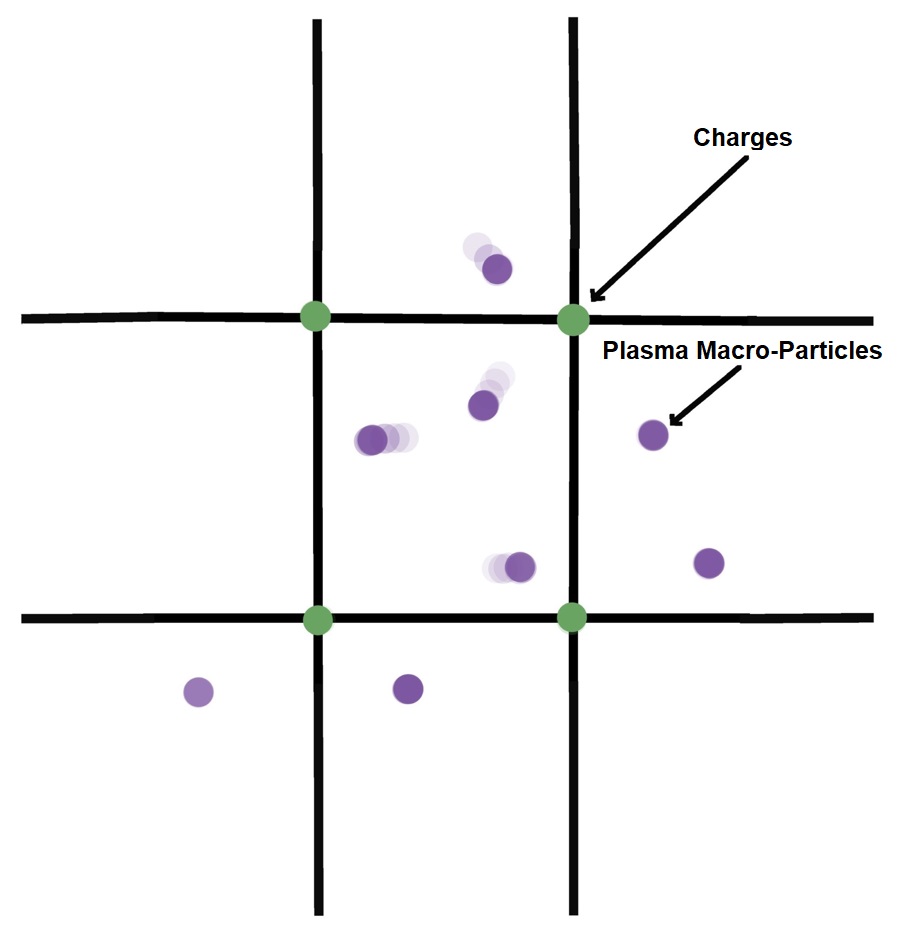}
    \caption{Particle-in-cell}
    \label{fig:pic}
  \end{subfigure}
  \caption{2D plasma physics simulation representations}
  \label{fig:bothfigures}
  %\vspace{-1.5em}
\end{figure}

%Plasma is formed when matter in the gaseous state is exposed to sufficiently high temperatures. 
%To form plasma, atoms are separated from some or the entirety of their electrons, thus transforming into ions that can be stimulated through electromagnetism to collide with each other and fuse. Because of plasma's property of quasineutrality, its possible to find local charge imbalances in large-scale plasma sample, but regardless of this, the net charge of the bulk plasma remains balanced with a value of zero. 
%A phenomenon related to quasineutrality that also occurs on large scale plasma is Debye shielding, whereby individual positively charged ions around them attracting negative ions and electrons forming a cloud around it that repulses other negatively charged particles producing a collective screening effect over the electric field produced by each individual particle over a certain length called the Debye length.

%\MJR{a heavier ! even lighter}
\vspace{-1.5em}
Fusion energy is generated in form of the released binding energy when light atomic nuclei (\textit{e.g.} Hydrogen) fuse to form a heavier nucleus (\textit{e.g.} Helium). This process is controlled in thermonuclear fusion devices such as stellarators, which consist of a helically symmetric toroidal tube encircled by complex solenoidal coils that produce a strong magnetic field which confines the plasma within the vessel. The design, assembly, and actual experimental testing of a stellarator is technologically complex and expensive to carry out. Therefore, different scenarios and configurations are tested and optimized in simulations before they are actually manufactured~\cite{miloch2014plasma,picot2021magnetic}.

%\subsubsection{Field-line Tracer}
\noindent\textbf{Field-line tracer}.
% Numerical methods such as the fourth-order Runge-Kutta method are commonly employed for this purpose
%Magnetic fields can be represented and traced as a series of field lines that plasma particles would then use as a path under the influence of the Lorentz force exerted by the magnetic field.
Magnetic fields can be represented and traced as a series of field-lines that exert a Lorentz force on charged particles in a plasma (visualized in Figure~\ref{fig:stellarator}), to analyze and gain insights into how the magnetic field influences their trajectories. This is usually done by approximating the differential equations that describe the behavior of the magnetic field in the simulated space. This simulation is particularly useful for studying spaces like those used in plasma confinement, such as stellarators and tokamaks. %Equation~\ref{eq:flt} depicts the equations that describe a magnetic field line, represented as a vector function \textbf{r}(s), with \textbf{B} representing the magnetic field, and $s$ being a parametric curve on \textbf{r}. It also illustrates the formula for obtaining the length of a line segment.

\begin{comment}

\begin{equation}
\resizebox{0.9\linewidth}{!}{$
\textit{d } \mathbf{r} \parallel \mathbf{B},
\quad\frac{\textit{{d} }\mathbf{r}}{\textit{d} s} = \frac{\mathbf{B}}{B},
\quad\mathbf{r}(s_1) \rightarrow \mathbf{r}(s_2) = \left| s_2 - s_1 \right| \quad \text{for any } s_1, s_2
$}
\label{eq:flt}
\end{equation}
\end{comment}

%\subsubsection
%\noindent\textbf{Particle-in-cell}
%PIC, short for Particle-In-Cell, is a simulation technique used to study plasma behavior. In this method, plasma is represented as a collection ``macro-particles" that move freely on a grid structure represented in figure \ref{fig:pic}. Each grid point contains fixed charges, creating an electrostatic field that exerts forces on the particles, causing them to accelerate towards specific directions.
%During each iteration of the simulation, the particle movements are updated, and the charge densities are recalculated to account for the particles' influence as they move between grid cells. These updated charge densities, are then used to solve Maxwell's equations and calculate the electric field at each grid point. Finally, the electric field is evaluated at the locations of each particle, preparing the simulation for the next iteration.
%The differential equations solved via this method are a class of unsteady equations common in plasma physics.\cite{miloch2014plasma}\cite{Georganas2016}

%\subsubsection
\noindent\textbf{Particle-in-cell}. In this numerical method (\textit{cf.} Figure~\ref{fig:pic}), plasma is represented as a collection of charged ``macroparticles" of different kinds (typically ions and electrons) within a grid structure. The grid is used to numerically solve Maxwell's equations \cite{jackson_classical_1999}, given the electric charge and current densities obtained by interpolation from the positions of the particles. The latter, in turn, are evolved in time by the acceleration computed from the (interpolated) electromagnetic forces, as obtained from the solution of the Maxwell equations~\cite{jackson_classical_1999,fleisch_2008,Georganas2016,miloch2014plasma}.
%The differential equations solved by this method are a class of unsteady equations common in plasma physics.

%\vspace{-1.5em}
% \begin{equation}
%     \nabla\cdot \textbf{E} = \frac{\rho}{\epsilon_0}
%     \label{eq:electricflux}
% \end{equation}
% \begin{equation}
%     \nabla\cdot \textbf{B} = 0
%     \label{eq:magneticflux}
% \end{equation}
% \begin{equation}
%     \nabla\times\textbf{E} = -\frac{\partial\textbf{B}}{\partial t}
%     \label{eq:Faraday}
% \end{equation}
% \begin{equation}
%     \nabla\times\textbf{B} = \mu_0\left(\textbf{j}+\epsilon_0\frac{\partial\textbf{E}}{\partial t}\right)
%     \label{eq:ampere}
% \end{equation}

%\begin{equation}
%    \resizebox{0.9\linewidth}{!}{$
%    \nabla\cdot \mathbf{E} = \frac{\rho}{\epsilon_0}, \quad \nabla\cdot \mathbf{B} = 0,\quad
%    \nabla\times\mathbf{E} = -\frac{\partial\mathbf{B}}{\partial t}, \quad \nabla\times\mathbf{B} = \mu_0(\mathbf{j}+%%%\epsilon_0\frac{\partial\mathbf{E}}{\partial t})
%    $}
%    \label{eq:maxwell}
%    %\vspace{-1.em}
%\end{equation}
\vspace{-.5em}
\subsection{Performance Portability}

The performance of an application can be measured by different metrics, for example, by the time-to-solution, the number of floating point operations per second (FLOP/s), or energy consumption, among others. In this paper, we adopt the definitions for performance, portability, and performance portability provided by Pennycook \textit{et al.}~\cite{pennycook2016metric}. \emph{Performance} is ``any measurable property of an application's correct execution of a problem on a platform.'' \emph{Portability} is ``the ability of an application to execute a problem correctly on a given set of platforms''. Finally, \emph{performance portability} is ``a measurement of an application's performance efficiency for a given problem that can be executed correctly on all platforms in a given set''. For our different experiment configurations (platform, simulation, and programming model combination), we shall measure efficiency as their \emph{application efficiency}, which is defined as the ``achieved performance as a fraction of their best observed performance'', for example, the best achieved time-to-solution on a certain platform and simulation combination. Additionally, we use their proposed metric of performance portability \textipa{\textqplig}, which is given by the harmonic mean of $i$ experiment configurations' performance efficiency $e$ observed across a set of platforms $H$ (in our case, HPC systems) when running our applications $a$ (for this study, a programming model) that solve a problem $p$ (one of the simulations methods). Note that if a certain experiment configuration is not available for a certain platform in $H$, performance portability is nil. This is given by Equation~\ref{performance-portability-eq}.
%\MJR{what does "different simulations" mean here? same code and setup on different platforms/processors?, different programming models?}
%\MJR{see above: "best" of what?}
%\MJR{too unspecific: each simulation}

\begin{equation}
\text{\textipa{\textqplig}}(a, p, H) = 
\begin{cases}
    \frac{|H|}{\sum_{i \in H} \frac{1}{e_i(a, p)}} & \text{if } i \text{ is supported } \forall\ i \in H \\
    0 & \text{otherwise}
\end{cases}
\label{performance-portability-eq}
\end{equation}
%\MJR{what does" if i is supported, forall" actually mean ?}
%\textqplig

%Since different computers are designed to perform better in different areas, it is important to evaluate the performance measurement by comparing it with the performance that the platform is theoretically capable of.~\cite{PENNYCOOK2019947}.
%The Roofline performance model has also been adopted as a way to measure performance in different architectures.~\cite{8639946}
%An application can be referred to as \emph{performance portable} when it can be run efficiently on the different platforms that the application should run on.~\cite{9484790} 

Different libraries or programming models have been developed to program performance portable code. We chose Kokkos to adapt our simulations because of the advantages it provides as a popular, well-documented, well-maintained, high-level parallel programming library meant for shared-memory systems.

%it is meant for parallel algorithms executed on many-core chips that share memory, which matches the structures on which our experiments were run. 
%Kokkos allows programmers to execute code on both the GPU and CPU without the need for extensive code refactoring. 

%As we are working with pre-existing code in this study, Kokkos was a good choice to study the developer experience of porting code from one library to another, given that Kokkos is  documented, well maintained, .
%\MJR{we need to elaborate a bit more here !}
% as suggested in the related work? 

% Kokkos, stencil example, RAJA
%\vspace{-1.em}
\vspace{-.75em}
\subsection{Related Work}

A number of previous studies in performance portability have employed the metric we defined in Equation~\ref{performance-portability-eq} as a standard~\cite{1118945642,2229652861,3Deakin2020,4448639933,5558639943,8945642,7muralikrishnan2022scaling,6WRIGHT2024109123}. Deakin\textit{ et al.}~\cite{1118945642} obtained their best performance portability results using OpenMP and Kokkos in one of the largest studies of performance portability across different applications, architectures and programming models, on which they note that some overhead should be expected given that Kokkos' CPU abstraction is implemented using OpenMP. Deakin \textit{et al}.~\cite{3Deakin2020} also report their best performance portability results when using Kokkos and OpenMP over various applications and architectures. Kwack \textit{et al}.~\cite{8945642} employed a variety of performance portability metrics to measure various programming models on different architectures across different applications. Their findings highlight that each model has distinct strengths and the optimal choice of model varies depending on the specific application. Muralikrishnan \textit{et al.}~\cite{muralikrishnan2022scaling} investigated performance portability and scalability in a series of mini-apps related to the PIC scheme, on which Kokkos structures were used with OpenMP and MPI.

%Kwack et al.~\cite{8945642} focused on exploring the use of various programming models on different architectures and measure the performance portability achieved across different applications using a variety of metrics

Similar observations are noted by Wright \textit{et al}.~\cite{6WRIGHT2024109123} in their performance portability survey of plasma physics simulations. The findings from these studies collectively indicate that both our chosen performance portability metric and programming model (out of many more options, \textit{e.g.} RAJA~\cite{RAJA}, SYCL~\cite{SYCL}, Alpaka~\cite{alpaka}, StarPU~\cite{starpu}, among others) are reasonable and timely choices.

% implementation
%\vspace{-1.em}
\vspace{-1.25em}
\section{Code Implementation}
%\MJR{find a better heading, we do not really "design" anything here, maybe: "Code implementations"}
%\vspace{-0.35em}
\vspace{-.5em}
\subsection{Baseline Implementation}
%For our implementations, we assume that while our code could be further improved, its programmed in the way an experienced developer would.
%\subsubsection{Field-line Tracer}

\noindent\textbf{Field-line Tracer}.
The Biot-Savart Solver for Computing and Tracing Magnetic Field-Lines (BS-SOLCTRA)~\cite{Jimenez2019BS-SOLCTRA} has been used to study plasma confinement in fusion devices. The simulation applies the field-line tracing technique in a 3D vacuum magnetic field and uses Biot-Savart's law to test distinct modular electromagnetic coil configurations. %that would lead to a successful plasma particle confinement or a particle divergence threshold. 

The movement of each line, represented by a computational particle, is obtained by updating its current coordinates in each iteration of the simulation by computing the effect of the magnetic field that the set of modular coils generates, ignoring any impact by the interaction between particles. The integration employs a fourth-order Runge-Kutta method. This code was chosen due to its familiarity to the collaborating laboratories and its significance in their research.

The baseline code employed OpenMP to offload computations to the GPU, enabling the calculation of particle positions in parallel through shared memory. This code stores particles as one-dimensional arrays to better fit the GPU architectures, having been previously adapted from a CPU-only implementation that stored particles as an AoS and used SIMD-vectorization to optimize performance. A straightforward MPI implementation let us study the simulation's behaviour when distributing the particles between different computer nodes and GPUs. The performance of BS-SOLCTRA is measured in terms of time-to-solution, with lower values indicating higher application performance.
% We started with the baseline implementation to remove OpenMP and integrate Kokkos.

\noindent\textbf{Particle-in-cell}. The Parallel Research Kernels (PRK)~\cite{VanderWijngaart2014} is a suite of computational kernels related to linear algebra, stencil computations, graph algorithms, and particle simulations, among others. PRKs are designed to be tested in an exploratory manner with different HPC programming models, such as Kokkos, and OpenMP. One kernel of particular interest to this study is a PIC code that uses Coulomb's law to calculate the forces exerted on each particle by the electromagnetic fields generated by the grid charges of its current cell. To induce load imbalance and control the evolution of the particles' distribution, they can be initialized using four different patterns. 

%, and then we used the same Kokkos code to run the simulation on a GPU instead
The baseline code uses OpenMP on CPU, and we used it here to incorporate Kokkos. The computational performance of PIC is conventionally measured as the number of particles moved per second, a metric which we adopt in the following, with larger values indicating higher application performance.

Note that the adaptations of the two codes employed in this study were developed in opposite directions, \textit{i.e.},~BS-SOLCTRA was ported from GPU to CPU, and the other way around with PIC. \footnote{The code variants used can be found in this version control repository: https://gitlab.com/CNCA\_CeNAT/MPCDF/kokkos-bs-solctra.} 

\vspace{-1.5em}
\subsection{Kokkos Adaptation}
\vspace{-.5em}
Integration of Kokkos into the base codes required converting the data structures used to store the information about the particles, coils, among others, from C-style arrays of elements to Kokkos views. The conversion involved updating the source and header files to accommodate the new data structures, as certain functions in the modularized source code files relied on passing these data structures as parameters. Listing~\ref{data} exemplifies how data is handled in the different variants, and Listing~\ref{parallel} shows how the computation kernels were parallelized. 

\begin{lstlisting}[numbers=none,caption=Data Management.,language=C++,label=data,basicstyle=\fontsize{7}{7.5}\selectfont\ttfamily, lineskip=0pt  ]
// OpenMP
// Memory allocation
particles = static_cast<double*>(aligned_alloc(alignment_size, sizeof(double)*particles_size));
// Initialize data normally...
// Device data transfer
#pragma omp target enter data map(to:particles[0:particles_size])
// --------
// Kokkos
// Memory space setting and memory allocation
#ifdef KOKKOS_ENABLE_CUDA
#define MemSpace Kokkos::CudaSpace
#endif
#ifdef KOKKOS_ENABLE_HIP
#define MemSpace Kokkos::HIPSpace
#endif
#ifndef MemSpace
#define MemSpace Kokkos::HostSpace
#endif

typedef Kokkos::View<double*, MemSpace> ViewVectorType;
ViewVectorType particles("particles", particles_size);
ViewVectorType::HostMirror h_particles = Kokkos::create_mirror_view(particles);
// Initialize data normally...
// Device data transfer
Kokkos::deep_copy(particles, h_particles);
\end{lstlisting}

\begin{lstlisting}[numbers=none,caption=Kernel Parallelization.,language=C++,label=parallel,basicstyle=\fontsize{7}{7.5}\selectfont\ttfamily, lineskip=0pt  ]
// OpenMP 
#pragma omp parallel num_threads(2)
{
    #pragma omp single
    {
    for (int i = 1; i <= steps; i++){
        #pragma omp target teams distribute parallel for
        for(int p=0; p < particle_count ; p++){
            // simulation computation...
    }}}}
// --------
// Kokkos
// Execution space and range policy setting
using ExecSpace = MemSpace::execution_space;
using Kokkos::RangePolicy<ExecSpace> range_policy;
Kokkos::parallel_for("run_particles_parallel_reduce", range_policy(0,particle_count), KOKKOS_LAMBDA (int p)
    {
        // simulation computation... 
    });

\end{lstlisting}

In the PIC OpenMP code, the particles are defined as \texttt{structs} and then dynamically allocated in an array using pointer arithmetic, while the grid structure is created using traditional \texttt{malloc} calls. To integrate Kokkos, these data structures were adapted by keeping the \texttt{struct} definition and storing it within a Kokkos view, and replacing the \texttt{malloc} with a Kokkos view allocation. Finally, keywords like \texttt{KOKKOS\_ARCH = "Ampere80"} are used so that Kokkos can hint the compiler to optimize the code for a specific target architecture. %\MJR{which variables? the Makefile organization is not relevant here, but we should describe the relevant compile-time configuration a bit more explicitly}

\begin{comment}
Parallel dispatch through Kokkos was achieved with parallel operations as displayed in Listing~\ref{parallel-dispatch}.

\begin{lstlisting}[numbers=none, caption=Kokkos parallel dispatch over particles.,label={parallel-dispatch}]
Kokkos::parallel_for("update_xyz", range_policy(0,particle_count), KOKKOS_LAMBDA (int p) {
    // for multidimensional particle coordinates
    base = p*DIMENSIONS;
    computeIteration(&particles(base));
});
\end{lstlisting}

\end{comment}

% results
%\vspace{-1.em}
\vspace{-1.5em}
\section{Experiments}
\vspace{-.5em}
\subsection{Setup}
%Distinct partitions were used, with some simulations relying on CPU computation, and others being GPU-accelerated.
The simulations for this study were conducted on four different HPC systems: Kabr\'e at Costa Rica's National High Technology Center (CeNAT), Raven at the Max Planck Computing and Data Facility (MPCDF), and benchmark systems at AMD and Nvidia, respectively. 
%~\cite{Raven} ~\cite{Kabre}
The specifications of the machines and the corresponding software configuration can be found in Table~\ref{hpc-systems} and Table~\ref{software-config-table}, respectively.
Note that our GPU codes do not make use of the novel ``unified" memory architecture connecting the GPU and the CPU parts of the latest Nvidia GH200 and AMD MI300A chips, \textit{i.e.},~for our benchmarks, only the GPU parts of these platforms are relevant. 

% \usepackage{multirow}

\vspace{-2.em}
\begin{table}[]

\caption{HPC Systems: hardware configuration}
\resizebox{\textwidth}{!}{%
\begin{tabular}{|c|c|c|c|c|c|c|c|c|}
\hline
\textbf{Machine}                       & \textbf{Partition} & \textbf{Nodes} & \textbf{CPU}              & \textbf{\begin{tabular}[c]{@{}c@{}}Cores\\ per node\end{tabular}} & \textbf{SMT} & \textbf{GPU}             & \textbf{\begin{tabular}[c]{@{}c@{}}RAM (GB)\\per node\end{tabular}} & \textbf{\begin{tabular}[c]{@{}c@{}}Theoretical\\FP64 peak TFlop/s\\per node\end{tabular}} \\ \hline
\multirow{3}{*}{Kabré}                 & CPU                & 8              & {\begin{tabular}[c]{@{}c@{}}Intel Xeon\\Gold 6354\end{tabular}}      & 36                      & yes          & -                        & 512                        & 1.9*                                          \\ \cline{2-9} 
                                       & GPU                & 4              & {\begin{tabular}[c]{@{}c@{}}Intel Xeon\\Silver 4214R\end{tabular}}   & 24                      & -            & NVIDIA V100              & 32                         & 7                                               \\ \hline
\multirow{3}{*}{Raven}                 & CPU                & 1592           & $\quad${\begin{tabular}[c]{@{}c@{}}Intel Xeon\\Platinum 8360Y\end{tabular}}$\quad$ & 72                      & yes          & -                        & 256                        & 3*                                              \\ \cline{2-9} 
                                       & GPU                & 192            & {\begin{tabular}[c]{@{}c@{}}Intel Xeon\\Platinum 8360Y\end{tabular}} & 72                      & yes          & $\quad${\begin{tabular}[c]{@{}c@{}}NVIDIA\\A100-SXM4-40GB\end{tabular}}$\quad$  & 512                        & 9.7                                             \\ \hline
\multirow{3}{*}{$\quad$AMD Accelerator Cloud$\quad$} & GPU                & 1              & {\begin{tabular}[c]{@{}c@{}}2x AMD\\EPYC 7F52\end{tabular}}          & 2x 16                   & -            & AMD MI210                & 512                        & 22.6                                            \\ \cline{2-9} 
                                       & GPU                & 1              & AMD Zen 4                 & 24                      & yes          & {\begin{tabular}[c]{@{}c@{}}AMD MI300A\\(A0 stepping)\end{tabular}} & 128                        & 81.7                                            \\ \hline
NVIDIA GH200                           & GPU                & 1              & NVIDIA Grace              & 72                      & -            & NVIDIA H100              & 480                        & 34                                              \\ \hline
\multicolumn{9}{l}{\footnotesize{* Reported by Microway~\cite{microway_icelake}}}
\end{tabular}}
\label{hpc-systems}
%\tablefootnote{*}{hola}
%\footnote{This is the footnote explaining the table.}hol.a
%\footnotetext{}{\footnotesize{* Reported by Microway~\cite{microway}}}
\end{table}

\begin{table}[]
%\vspace{-1.5em}
\caption{HPC Systems: software configuration}
\centering
%\footnotesize
\resizebox{\textwidth}{!}{%
%\scalebox{0.65}{ %0.575
\begin{tabular}{|c|c|c|c|c|}
\hline
\textbf{Machine} & \textbf{OS} & \textbf{Compiler} & \textbf{Kokkos} & \textbf{MPI} \\ \hline
Kabré            & CentOS 7    & $\quad$GCC 9.2 + CUDA 11.6.2 $\quad$  & 4.3.0-RC           & -  \\ \hline
Raven            & $\quad$SUSE Linux Enterprise Server 15-SP3$\quad$            & NVIDIA HPC-SDK 22                     & 4.3.0-RC           & $\quad$Open MPI 4$\quad$     \\ \hline 
$\quad$AMD Accelerator Cloud$\quad$            & Ubuntu 22.04.2 LTS     & ROCm 6.1.0              & $\quad$4.3.0-RC$\quad$           & -  \\ \hline
NVIDIA GH200            & Rocky Linux 9.3     & NVIDIA HPC-SDK 24.3              & 4.3.0-RC          & -  \\ \hline
\end{tabular}}
\label{software-config-table}
%\vspace{-3em} % Reduce space after the table
\vspace{-1.5em}
\end{table}
%\vspace{-0.4cm} 

\vspace{-1.5em}
Computational performance for BS-SOLCTRA was measured for three different problem sizes over 1,000 iterations of the simulation, using the same setup as in our previous study~\cite{10.1007/978-3-031-23821-5_3}. For PIC, four particle initialization patterns were used to induce different load imbalance scenarios. The problem sizes and the initialization patterns are stated in Table~\ref{experiment-configs}. The performance metric is time-to-solution in seconds (lower is better) for BS-SOLCTRA, and the rate of particles moved per second (higher is better) for PIC. For each measurement we report the arithmetic mean of 10 executions, with no notable run-to-run variations.

\vspace{-0.8cm}
\begin{table}[]
\caption{Simulation setup}
\resizebox{\textwidth}{!}{%
%\scalebox{0.55}{
\begin{tabular}{|c|c|c|c|c|c|}
\hline
\textbf{Simulation} & \multicolumn{1}{l|}{\textbf{Architecture}} & \textbf{Processing unit} & \textbf{Code variant} & \textbf{Problem size} & \textbf{Initialization pattern} \\ \hline
\multirow{3}{*}{$\quad$BS-SOLCTRA$\quad$} & CPU & Xeon Platinum 8360Y & $\quad$\multirow{5}{*}{\begin{tabular}[c]{@{}c@{}}Kokkos, \\ OpenMP\end{tabular}}$\quad$ & \multirow{3}{*}{\begin{tabular}[c]{@{}c@{}}102400,  512000,\\ 1024000 particles\end{tabular}} & \multirow{3}{*}{Random} \\ \cline{2-3}
 & GPU & $\quad$V100, A100, MI210, MI300A, H100$\quad$ &  &  &  \\ \cline{2-3}
 & GPU + MPI & A100* &  &  &  \\ \cline{1-3} \cline{5-6} 
\multirow{2}{*}{PIC} & CPU & \begin{tabular}[c]{@{}c@{}}Xeon Gold 6354,\\ Xeon Platinum 8360Y\end{tabular} &  & \multirow{2}{*}{$\quad$1000000 particles$\quad$} & \multirow{2}{*}{\begin{tabular}[c]{@{}c@{}}Geometric, Linear, \\ Patch, Sinusoidal\end{tabular}} \\ \cline{2-3}
 & GPU & V100, A100 &  &  &  \\ \hline
 \multicolumn{6}{l}{\footnotesize{* For our multi-node scaling experiments, 2, 4, 8 \& 16 A100 GPUs were used}} \\ %\hline |l| 
\end{tabular}}
\label{experiment-configs}
%\vspace{-3.em}
\end{table}

\vspace{-1.2cm}
\subsection{Field-line Tracer}
%\MJR{the heading doesn't really fit here, as we report not only baseline performance but also about the ported versions}
%\MJR{and how reasonable are our "baseline" performances actually?}

\begin{figure}
%\vspace{-1em}
  \centering
  \begin{subfigure}[b]{0.45\textwidth}
    \includegraphics[width=\textwidth]{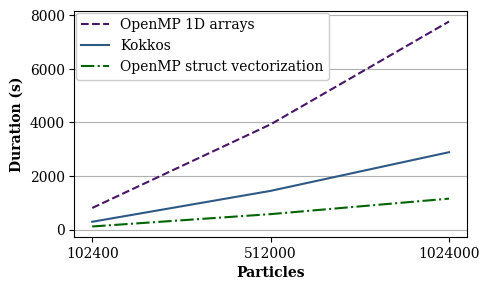}
    \captionsetup{justification=centering}
    \caption{CPU adaptation on Xeon Platinum 8360Y with the OpenMP (dashed, and dashed-dotted lines) and Kokkos variants (solid lines).}
    \label{fig:bs-solctra-execution-time}
  \end{subfigure}
  \hfill
  \begin{subfigure}[b]{0.45\textwidth}
    \includegraphics[width=\textwidth]{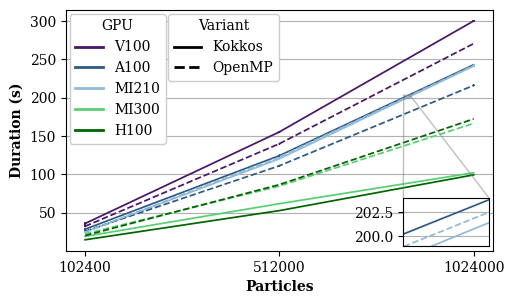}
    \captionsetup{justification=centering}
    \caption{GPU adaptation on different GPU models (colours) with the OpenMP variant (dashed lines) and the Kokkos variant (solid lines).}
    \label{fig:gpu-execution-time}
  \end{subfigure}
  \captionsetup{justification=centering}
  \caption{Time-to-solution for the three different problem sizes on a single CPU/GPU with the OpenMP and Kokkos code variants of BS-SOLCTRA.}
  \label{fig:bothBS-SOLCTRAs}
 \vspace{-1.5em}
\end{figure}

%The baseline implementation of BS-SOLCTRA is GPU-based.
% Performance drop due to Kokkos
\noindent\textbf{GPU results}. The Kokkos adaptation of BS-SOLCTRA code resulted in a performance drop for the older V100 and A100 GPUs compared to the original OpenMP-based code variant. On the MI210 GPU, Kokkos performed as well as OpenMP. However, for the very latest generation of H100 GPU and MI300A APU, OpenMP shows an apparent underperformance when compared with the Kokkos results (\textit{cf.}~Figure~\ref{fig:gpu-execution-time}).
This may hint at OpenMP compilers (contrary to the platform-native CUDA and ROCM backends employed by Kokkos) being not yet fully tuned for this very latest generation of high-end GPUs. For the MI300A, in addition, all our measured performance numbers have to be
interpreted with some caution, as we had only access to a pre-production sample ``A0 stepping") which does not yet meet all the final specifications. At the same time it is remarkable that Kokkos delivers very decent out-of-the-box performance
for hardware that was introduced only very recently. 
%These results also highlight, as discussed in the literature cited in Section 2, that there is no single combination of programming model %and computer architecture that is always optimal for all applications.
Not too unsurprisingly, and consistent with findings reported in the literature cited in Section 2, there is no single combination of programming model and computer architecture that is always optimal for all applications.

% arch comparison

When comparing BS-SOLCTRA's code variants across GPU architectures, the H100 and MI300A GPUs demonstrated a clear performance hierarchy, outperforming the previous-generation A100 and MI210, respectively. In turn, the A100 and MI210 outperformed the older V100 GPU. These results are broadly consistent across code variants and all GPU architectures (except for the case on which the A100 performed better than the MI210 with OpenMP, while they achieved similar performance with Kokkos), and reflect the compute-bound nature of our application in relation to the peak Tflop/s ratings of the different GPUs (\textit{cf.} Table~\ref{hpc-systems}). 

%The A100 outperformed the MI210 with a speedup of 1.12 with OpenMP, while with Kokkos, there was no significant difference. The newer NVIDIA H100 GPU demonstrated a speedup of 1.4 over the MI210 and 1.25 over the A100 with OpenMP. With Kokkos, this performance gap even widened, with the H100 achieving a speedup of 2.44 over the MI210 and the A100.

%When comparing BS-SOLCTRA's code variants across GPU architectures, the current Nvidia (A100) and AMD (MI210) GPUs, both outperformed the previous-generation V100 by relative factors of 1.24 and 1.21 with Kokkos, respectively. On the other hand, when using OpenMP, both GPUs yielded a speedup of 1.12 over the V100.
%The A100 outperformed the MI210 with a speedup of 1.12 with OpenMP, while with Kokkos, there was no significant difference. The newer NVIDIA H100 GPU demonstrated a speedup of 1.4 over the MI210 and 1.25 over the A100 with OpenMP. With Kokkos, this performance gap even widened, with the H100 achieving a speedup of 2.44 over the MI210 and the A100.

The different setups resulted in a near-linear relation of problem size with time-to-solution on the various GPUs tested. The smaller problem sizes likely failed to fully saturate the more powerful GPUs, which explains the comparably better performance of the V100 for the smallest problem size.
%and the H100 matching the MI300A results.

%% PP measurementes.
%To measure the performance portability of our simulation with the metric in Equation~\ref{performance-portability-eq}, we took our Kokkos and OpenMP code variants as the applications under evaluation. For each application, we selected the best performance achieved on each architecture for the largest problem size. We then identified the overall best performance across both applications to calculate application efficiencies shown in Table~\ref{eff-bs-solctra}. Considering all the GPUs used, the measured performance portability (Table~\ref{perf-port}) was \SI{96}{\percent} for our Kokkos implementation, and \SI{78}{\percent} for OpenMP.  
We used the metric in Equation~\ref{performance-portability-eq} to measure the performance portability of our simulation. For each application (code variant), we selected the best performance achieved on each architecture for the largest problem size. We then identified the overall best performance across both applications to calculate application efficiencies shown in Table~\ref{eff-bs-solctra}. Considering all the GPUs used, the measured performance portability (Table~\ref{perf-port}) was \SI{96}{\percent} for our Kokkos implementation, and \SI{78}{\percent} for OpenMP.  

\vspace{-.75cm}
\begin{table}
%\vspace{-2.75em}
\caption{BS-SOLCTRA application efficiency for the largest problem size (1024000 particles)}
%\resizebox{\textwidth}{!}{%
    \centering
\scalebox{0.75}{
\begin{tabular}{|c|c|r|r|}
\hline
\textbf{Platform}     & \multicolumn{1}{l|}{\textbf{Code variant}} & \textbf{Time-to-solution} & \textbf{Application efficiency} \\ \hline
\multirow{2}{*}{V100} & OpenMP   & 270.39  & \SI{100}{\percent} \\ \cline{2-4} & Kokkos & 300.08 & \SI{90}{\percent} \\ \hline
\multirow{2}{*}{A100} & OpenMP & 216.05 & \SI{100}{\percent} \\ \cline{2-4} & Kokkos & 242.91 & \SI{89}{\percent} \\ \hline
\multirow{2}{*}{MI210}& OpenMP & 241.53 & \SI{99}{\percent} \\ \cline{2-4} & Kokkos & 241.35 & \SI{100}{\percent} \\ \hline
\multirow{2}{*}{MI300A} & OpenMP & 166.52 & \SI{61}{\percent} \\ \cline{2-4} & Kokkos & 102.09 & \SI{100}{\percent} \\ \hline
\multirow{2}{*}{H100} & OpenMP & 172.87 & \SI{57}{\percent} \\ \cline{2-4} & Kokkos & 98.95 & \SI{100}{\percent} \\ \hline
\multirow{2}{*}{\begin{tabular}[c]{@{}c@{}}Xeon Platinum \\ 8630Y\end{tabular}} & OpenMP & 7763.10   & \SI{37}{\percent} \\ \cline{2-4} & Kokkos                      & 2889.40 & \SI{100}{\percent} \\ \hline
\end{tabular}}
\label{eff-bs-solctra}
\vspace{-1.em}
\end{table}

% 99% vs 100%

\noindent\textbf{CPU results}.
%Using an Intel Xeon Platinum 8360Y CPU node, we compare our Kokkos adaptation of BS-SOLCTRA with the SoA and AoS OpenMP variants.\footnote{Note that in our previous paper~\cite{10.1007/978-3-031-23821-5_3}, where we ported our OpenMP CPU code to the GPU, using a SoA memory layout lead to better performance in the GPU, while the use of AoS was better for the CPU. For this reason, our AoS OpenMP code variant was not ported to the GPU.} The AoS code variant applies AVX-512 instructions to provide SIMD vectorization support through SIMD language directives and compiler flags \textit{(-mavx512f -mavx512pf -mavx512er -mavx512cd)} to perform the magnetic field related computations in parallel to enhance CPU performance. The impact of this enhancement can be observed in Figure~\ref{fig:bs-solctra-execution-time}.
Using an Intel Xeon Platinum 8360Y CPU node, we compare our Kokkos adaptation of BS-SOLCTRA with the one-dimensional array and AoS OpenMP variants.\footnote{In our prior work~\cite{10.1007/978-3-031-23821-5_3}, porting the OpenMP CPU code variant to the GPU favored one-dimensional arrays for performance, while AoS remained preferable for the CPU. For this reason, our AoS OpenMP code variant was not ported to the GPU.} The AoS code variant relies on (guided) autovectorization by the compiler, \textit{i.e.}~we apply the \texttt{\#pragma omp simd} directive, together with the relevant compiler flags (\texttt{-mavx512f -mavx512pf -mavx512er -mavx512cd}) in order to enable AVX-512 instructions for the Intel Xeon CPUs. The impact of using the different code variants can be observed in Figure~\ref{fig:bs-solctra-execution-time}.

\begin{comment}
  \begin{figure}
  \centering
    \includesvg[width=0.5\textwidth]{./figures/BS-SOLCTRA-CPU.svg}
    \captionsetup{justification=centering}
    \caption{Time-to-solution for the three different problem sizes on a single Intel Xeon Platinum 8360Y Node with the two OpenMP variants (dashed, and dashed-dotted lines) and the Kokkos variant (solid lines) of BS-SOLCTRA.}
    \label{fig:bs-solctra-execution-time}
  \end{figure}
\end{comment} 

%\MJR{throughout the paper: be consistent with using present or tense past tense etc. !}
% with its CPU memory space data allocation
%\MJR{the following two paragraphs are unclear and imprecise, rewrite !}
Our Kokkos implementation achieved a 2.7x speedup compared to the sub-optimal OpenMP one-dimensional array variant, demonstrating portability benefits. However, it fell short of the performance of the AoS OpenMP variant. While Kokkos exhibited a 2.5x slowdown relative to AoS OpenMP, this is significantly less severe than the 6.7x slowdown observed with one-dimensional array OpenMP. When included in the set of platforms with the GPUs previously used, Kokkos achieved an application efficiency of \SI{100}{\percent} and a performance portability of \SI{96}{\percent}. By contrast, the one-dimensional array OpenMP variant's application efficiency of \SI{37}{\percent} resulted in a performance portability of \SI{66}{\percent}. This comparison demonstrates Kokkos's ability to deliver more consistent performance across diverse architectures.~\footnote{In this measurement, we intentionally used our sub-optimal one-dimensional array OpenMP variant for consistency with the code employed in our GPU runs. The CPU optimized SoA code variant was included to provide a perspective of the impact in loss of performance in the architecture for which the code was originally optimized by programming performance portable code.} 

%Note that we used our sub-optimal one-dimensional array OpenMP variant in this measurement because that's the code that we used in our GPU runs as well. Our optimal baseline code variant was included to provide a perspective of the impact (The loss in performance in the architecture for which the coda had been specifically optimized) of programming performance portable codes.}   

\begin{figure}
%\vspace{-1em}
  \centering
  \begin{subfigure}[b]{0.45\textwidth}
    \includegraphics[width=\textwidth]{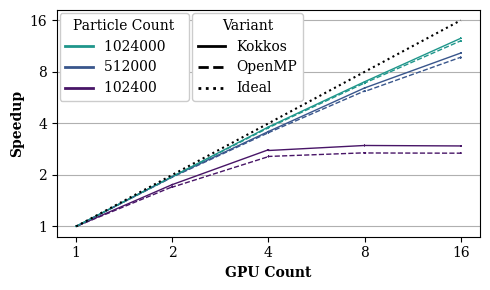}
    \captionsetup{justification=centering}
    \caption{Strong Scaling}
    \label{fig:bs-solctra-ss}
  \end{subfigure}
  \hfill
  \begin{subfigure}[b]{0.45\textwidth}
  %\includesvg[]{figures/BS-Solctra-WS.svg}
    \includegraphics[width=\textwidth]{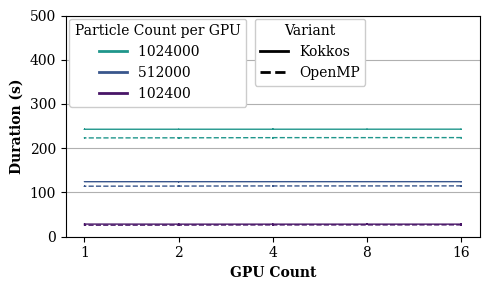}
    \captionsetup{justification=centering}
    \caption{Weak Scaling}
    \label{fig:bs-solctra-ws}
  \end{subfigure}
  \captionsetup{justification=centering}
  \caption{BS-SOLCTRA parallel scaling (runtime as a function of employed A100 GPUs) for the Kokkos (solid lines) and the OpenMP (dashed lines) code variants for three different problem sizes (colour coded).}
  \label{fig:bothfigures3}
 % \vspace{-1.5em}
\end{figure}

%\vspace{-0.25em}
\noindent\textbf{Multi-node scalability}. 
For completeness, Figure~\ref{fig:bs-solctra-ss} presents strong and weak scaling results on up to 16 A100 GPUs (4 Raven nodes). 
We note that the code does not involve particle interactions which would introduce communication or synchronization across GPUs and
hence ideal parallel scaling should be expected. While this is experimentally confirmed for the weak scaling (Figure~\ref{fig:bs-solctra-ws}), the strong scaling speedups from 1 to 16 GPUs are only 12.6 (Kokkos) and 12.1 (OpenMP). This can be explained by a progressive under-utilization of the individual GPUs as the workload per GPU decreases with increasing number of GPUs. Such under-saturation can also explain the performance difference between the Kokkos and OpenMP variants narrowing as more GPUs are used.
Overall, measured multi-node performances are consistent with the single-GPU findings discussed earlier, and no unexpected interference of the intra-node performance with inter-node MPI parallelization is found.

% arXiv
\begin{comment}
\vspace{-0.3cm}
\begin{figure}
%\vspace{-1em}
  \centering
  \begin{subfigure}[b]{0.45\textwidth}
    \includesvg[width=\textwidth]{./figures/BS-Solctra-SS.svg}
    \captionsetup{justification=centering}
    \caption{Strong Scaling}
    \label{fig:bs-solctra-ss}
  \end{subfigure}
  \hfill
  \begin{subfigure}[b]{0.45\textwidth}
  %\includesvg[]{figures/BS-Solctra-WS.svg}
    \includesvg[width=\textwidth]{./figures/BS-Solctra-WS.svg}
    \captionsetup{justification=centering}
    \caption{Weak Scaling}
    \label{fig:bs-solctra-ws}
  \end{subfigure}
  \captionsetup{justification=centering}
  \caption{BS-SOLCTRA parallel scaling (runtime as a function of employed A100 GPUs) for the Kokkos (solid lines) and the OpenMP (dashed lines) code variants for three different problem sizes (colour coded).}
  \label{fig:bothfigures3}
 % \vspace{-1.5em}
\end{figure}
\vspace{-0.7cm}
\end{comment}

\vspace{-0.3cm}
\vspace{-0.7cm}

%For each setup, the OpenMP variant outperformed its Kokkos counterpart, consistent with the single-GPU findings discussed earlier. 

%The increasing number of GPUs likely lead to under-utilization of the individual GPUs. This under-saturation could explain why the performance difference between Kokkos and OpenMP narrowed as more GPUs were added. Regardless of this, both programming models provided similar results with memory distribution. The Kokkos and OpenMP variants achieved strong-scaling speedups of 12.56 and 12.14, respectively, when comparing 16 GPUs to a single GPU. This behaviour was expected considering that MPI is independent from Kokkos and OpenMP and shouldn't cause a distinct impact on any of them.

%Our weak scaling results exhibit close-to-perfect application efficiency across all tested configurations (see Figure~\ref{fig:bs-solctra-ws}). This is expected because our simulation doesn't involve particle interactions that would introduce communication bottlenecks between GPUs.%\MJR{this argument is certainly plausible, but then I'm wondering why the strong-scaling is not ideal, too ?} 

%\vspace{-1.5em}
\subsection{Particle-in-cell}
\begin{figure}
%\vspace{-1em}
  \centering
  \begin{subfigure}[b]{0.45\textwidth}
    \includegraphics[width=\textwidth]{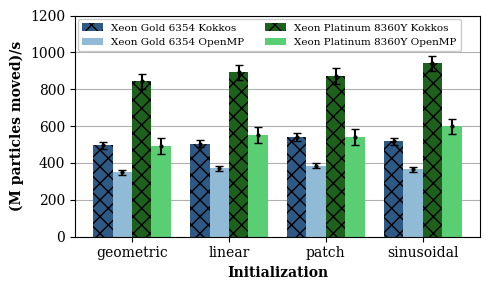}
    \captionsetup{justification=centering}
    \caption{{\parbox{\linewidth}{\centering{CPU adaptation on Xeon Gold 6354 (blue) and Platinum 8360Y (green).}}}}
    \label{fig:pic-dribe}
  \end{subfigure}
  \hfill
  \begin{subfigure}[b]{0.45\textwidth}
  %\includesvg[]{figures/BS-Solctra-WS.svg}
    \includegraphics[width=\textwidth]{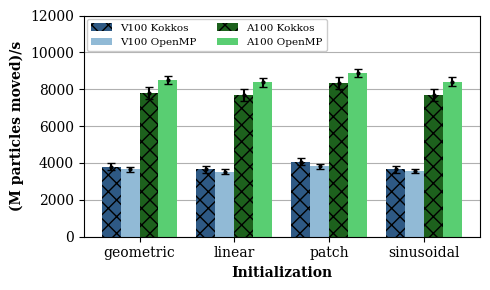}
    \captionsetup{justification=centering}
    \caption{{\parbox{\linewidth}{\centering{GPU adaptation on V100 (blue) and A100 (green).}}}}
    \label{fig:pic-initialization-rate-processing}
  \end{subfigure}
  \captionsetup{justification=centering}
  \caption{Performance on a single CPU/GPU for the Kokkos (cross hatch) and OpenMP (solid fill) variants of the PIC code for different initialization patterns. The black error bars indicate the measured run-to-run variations.}
  \label{fig:bothPICs}
 % \vspace{-1.5em}
\end{figure}

\noindent\textbf{CPU results}.
Figure~\ref{fig:pic-dribe} shows that, across all initialization patterns, our Kokkos adaptation outperformed the OpenMP variant in terms of particle movement rates. On Kabr\'e's Intel Xeon Gold 6354 CPU, Kokkos achieved a particle movement rate 1.4 times faster than OpenMP. This advantage increased to 1.6 times on Raven's Intel Xeon Platinum 8360Y CPUs. These results highlight the effectiveness of Kokkos' automatic memory allocation layout and its data access patterns optimizations, overcoming the need to implement architecture-specific data structures. This performance difference likely stems from the OpenMP variant storing the simulation grid in a one-dimensional array format, and considering the nature of GPU memory accessing patterns. Overall, the Xeon Platinum CPU reached particle movement rates about 1.7 times higher than the Xeon Gold for both code variants. 
% arch comparison

%\MJR{why do we thing SoA is bad here?}. %because of the nature of GPU memory access patterns.
%On the Xeon Gold CPUs, Kokkos achieved an increase of around 1.4X in particles moved per second compared to OpenMP. The difference became more significant when running the simulation on the Xeon Platinum CPUs, with Kokkos achieving a particle movement rate 1.6X higher than OpenMP

% arXiv change
\begin{comment}    
\begin{figure}
%\vspace{-1em}
  \centering
  \begin{subfigure}[b]{0.45\textwidth}
    \includesvg[width=\textwidth]{./figures/PIC-dribe.svg}
    \captionsetup{justification=centering}
    \caption{{\parbox{\linewidth}{\centering{CPU adaptation on Xeon Gold 6354 (blue) and Platinum 8360Y (green).}}}}
    \label{fig:pic-dribe}
  \end{subfigure}
  \hfill
  \begin{subfigure}[b]{0.45\textwidth}
  %\includesvg[]{figures/BS-Solctra-WS.svg}
    \includesvg[width=\textwidth]{./figures/PIC-GPU.svg}
    \captionsetup{justification=centering}
    \caption{{\parbox{\linewidth}{\centering{GPU adaptation on V100 (blue) and A100 (green).}}}}
    \label{fig:pic-initialization-rate-processing}
  \end{subfigure}
  \captionsetup{justification=centering}
  \caption{Performance on a single node for the Kokkos (cross hatch) and OpenMP (solid fill) variants of the PIC code for different initialization patterns. The black error bars indicate the measured run-to-run variations.}
  \label{fig:bothPICs}
 % \vspace{-1.5em}
\end{figure}
\end{comment}

%Since the different load-balancing scenarios didn't affect dthe performance by any significant amount\MJR{why? is there no actual imbalance in the particle distribution and/or was Kokkos and OpenMP effective in mitigating it dynamically? if yes how?}cause an impact in the behaviour of any experiment configuration

As the performance difference between configurations remained consistent across load-balancing scenarios, we arbitrarily chose the geometric initialization results to measure the application's performance portability. Based on the efficiencies in Table~\ref{eff-pic}, Kokkos achieved \SI{100}{\percent} performance portability, while OpenMP reached only \SI{67}{\percent} (Table~\ref{perf-port}).
%\MJR{shorten the rest of the sentence and instead reference the equation}, with the particle movement per second rate as our performance metric to measure our application efficiency against the best observation across all applications, as detailed in 
% \usepackage{multirow}

%0.9997514602
%\percentage{0.9276072341}

\begin{comment}
\begin{figure}
%\vspace{-1.5em}
    \centering
    %\vspace{-15pt} % Reduce space before the image
    \includesvg[width=.5\textwidth]{./figures/PIC-GPU.svg}
    \captionsetup{justification=centering}
    \caption{Performance on a single NVIDIA V100 (green) and A100 (blue) GPU node for the Kokkos (cross hatch) and OpenMP (solid fill) variants of the PIC code for different initialization patterns. The black error bars indicate the measured run-to-run variations.}
    \label{fig:pic-initialization-rate-processing}
    %\vspace{-100pt} % Reduce space after the image
%    \vspace{-1.5em}
\end{figure}
\end{comment}

%\subsubsection{PIC}
\noindent\textbf{GPU Results}. When running the simulation on a single V100 or A100 node, the particle movement rate, visualized in Figure~\ref{fig:pic-initialization-rate-processing}, increased by up to 13 times with Kokkos compared to the results previously obtained with the CPU implementation. OpenMP achieved an even higher improvement, reaching up to a 20 times increase in particle movement rate. This is expected for a compute-bound code like PIC, given the significant difference in Tflop/s between CPUs and GPUs.

\begin{table}
%\vspace{-2em}
\caption{PIC application efficiency for the geometric initialization pattern}
%\resizebox{\textwidth}{!}{%

  \centering
\scalebox{0.75}{
  
\begin{tabular}{|c|c|r|r|}
\hline
\textbf{Platform} & \multicolumn{1}{c|}{\textbf{Code variant}} &\textbf{\begin{tabular}[c]{@{}c@{}}Particle\\movement rate\end{tabular}} &  \textbf{\begin{tabular}[c]{@{}c@{}}Application\\efficiency\end{tabular}} \\ \hline

\multirow{2}{*}{V100} & OpenMP & 3665.62 &  \SI{96}{\percent} \\ \cline{2-4}  & Kokkos  & 3805.13 & \SI{100}{\percent} \\ \hline

\multirow{2}{*}{A100} & OpenMP & 8541.44 &  \SI{100}{\percent} \\ \cline{2-4}  & Kokkos  & 7867.13 & \SI{92}{\percent} \\ \hline

$\quad$\multirow{2}{*}{\begin{tabular}[c]{@{}c@{}}Xeon Platinum\\ 8630Y\end{tabular}}$\quad$ & $\quad$OpenMP$\quad$ & 612.49 & \SI{64}{\percent} \\ \cline{2-4} & Kokkos  & 960.67 & \SI{100}{\percent}  \\ \hline 

\multirow{2}{*}{\begin{tabular}[c]{@{}c@{}}Xeon Gold\\ 6354\end{tabular}} & OpenMP & 425.41 & \SI{72}{\percent}  \\ \cline{2-4} & Kokkos & 593.71 & \SI{100}{\percent} \\ \hline

\end{tabular}}
\label{eff-pic}
%\vspace{-4.5em}
\end{table}

\begin{table}%\fontsize{7}{9}
\caption{Performance portability measurements}
\centering
%\resizebox{8.5cm}{!}{%
\scalebox{0.8}{
\begin{tabular}{|c|c|c|r|}
\hline
\textbf{Simulation $p$}  & \textbf{Platform set $H$} & \textbf{Code variant $a$} & {\textipa{\textqplig}$(a, p, H)$}   \\ \hline 
 $\quad$\multirow{6}{*}{BS-SOLCTRA}$\quad$ & \multirow{2}{*}{\{V100, A100, MI210, MI300A, H100\}} & OpenMP & \SI{78}{\percent} \\ \cline{3-4} & & Kokkos & \SI{96}{\percent} \\ \cline{2-2} \cline{3-4} 

 & \multirow{2}{*}{\{Xeon Platinum 8630Y\}} & $\quad$OpenMP$\quad$ & \SI{37}{\percent}  \\ \cline{3-4} & & Kokkos & \SI{100}{\percent} \\ \cline{2-2} \cline{3-4} 

 & $\quad$\multirow{2}{*}{\begin{tabular}[c]{@{}c@{}}\{V100, A100, MI210, MI300A, H100, \\ Xeon Platinum 8630Y\}\end{tabular}}$\quad$ & OpenMP & \SI{66}{\percent} \\ \cline{3-4} & & Kokkos & \SI{96}{\percent} \\ \hline 
 
  \multirow{6}{*}{PIC} & \multirow{2}{*}{\begin{tabular}[c]{@{}c@{}}\{Xeon Platinum 8630Y,\\ Xeon Gold 6354\}\end{tabular}} & OpenMP & \SI{67}{\percent}  \\ \cline{3-4} & & Kokkos & \SI{100}{\percent} \\ \cline{2-2} \cline{3-4} 

& \multirow{2}{*}{\{V100, A100\}} & OpenMP & \SI{98}{\percent} \\ \cline{3-4} & & Kokkos & \SI{96}{\percent}   \\ \cline{2-4}

 & \multirow{2}{*}{\begin{tabular}[c]{@{}c@{}}\{V100, A100, Xeon Platinum 8630Y,\\  Xeon Gold 6354\}\end{tabular}} & OpenMP & \SI{80}{\percent} \\ \cline{3-4} & & Kokkos & \SI{98}{\percent} \\ \hline
\end{tabular}}
\label{perf-port}
%\vspace{-2.5em}
\end{table}

Our experiments revealed contrasting performance trends for the V100 and A100 GPUs. Regardless of the load balancing strategy, Kokkos outperformed OpenMP on the V100. However, the opposite was true for the A100, where OpenMP edged out Kokkos. The performance gap between Kokkos and OpenMP was narrower on both GPUs compared to the CPU versions (where Kokkos excelled). This suggests that Kokkos exhibits better performance portability across architectures. Overall, the A100 produced particle movement rates over 2 times higher than the V100, which is consistent with their Tflop/s difference.

%Running the simulaton on a Raven's A100 GPU, the particle movement rate, visualized in Figure~\ref{fig:pic-initialization-rate-processing}, increased by about 13 times with Kokkos in comparison with the results previously obtained with the CPU implementation regardless of the load balancing scenario. With OpenMP, the particle movement rate increase obtained was of about 15 times. This should be the case in this compute-bound simulation considering the difference in TFlop/s between the CPU and the GPU, and the fact that the grid data structure was not in its optimal layout for the CPU code. Overall, the A100 produced particle movement rates over 2 times higher than the V100, consisted with the 

The application efficiencies obtained for these runs were \SI{100}{\percent} for Kokkos and \SI{96}{\percent} for OpenMP with the V100, and \SI{92}{\percent} for Kokkos and \SI{100}{\percent} for OpenMP with the A100. The performance portability metric for the set of GPUs resulted in \SI{96}{\percent} for Kokkos and \SI{98}{\percent} for OpenMP. Integrating these GPU results with the set of CPUs used earlier, the performance portability metrics obtained were \SI{98}{\percent} for our Kokkos code and \SI{80}{\percent} for our OpenMP code.

\vspace{-1.em}
\subsection{Discussion}

%both in reducing time-to-solution and increasing the particle movement rates
%Kokkos provided ease of development, and also a simple way to optimize the code for specific processing unit models. This is given by the fact that minimal changes in the source code worked on the different architectures just by recompiling them with their specific architecture flag. Besides being a programmer friendly code experience, 

%Kokkos' polymorphic memory layouts proved to be very effective in enhancing performance for the simulations we studied when running them on different architectures. Kokkos performance portability measurements ranged between \percentage{0.9553} and \SI{100}{\percent} with a mean performance portability of \percentage{0.9759399594}, excelling OpenMP's metrics that ranged between \percentage{0.3722} and \percentage{0.9813} with a mean performance portability of \percentage{0.7122707139}.

Our experiments demonstrate that Kokkos' ability to adapt memory layouts across architectures significantly enhances performance portability. Kokkos achieved performance portability between \SI{96}{\percent} and \SI{100}{\percent}, with an average of \SI{98}{\percent}. This significantly outperforms OpenMP, whose portability ranged from \SI{37}{\percent} and \SI{98}{\percent}, with an average of \SI{71}{\percent}.
While Kokkos achieved good performance portability in BS-SOLCTRA (never falling below \SI{96}{\percent} and outperforming OpenMP's maximum of \SI{78}{\percent}), it did introduce a slowdown of up to 2.5 times in some configurations compared to the original, optimized code (which required significant architecture-specific refactoring). This slowdown might be acceptable considering the larger slowdowns observed with OpenMP (up to 6.7x with one-dimensional arrays on CPU). We believe further optimizations, particularly CPU-focused SIMD vectorization, could improve the performance of the Kokkos variant. Similarly, in PIC, Kokkos achieved performance portability measurements that never fell below \SI{96}{\percent} and surpassed OpenMP's in two out of three platform sets. While Kokkos may experience some slowdowns compared to optimized code, its portability benefits outweigh this drawback in many cases.

%\MJR{we haven't said anything, \textbf{how "easy"} it actually was for our codes. We should give at least a qualitative indication to the reader, who is basically interested in the question: do their findings transfer/apply to my own use-case ? THIS is the basic question that the paper actually has to address !}

%In BS-SOLCTRA, performance portability was achieved satisfactorily with Kokkos, never falling below \percentage{0.9553} and always measuring higher than OpenMP, that never measured over \percentage{0.7838}. Nonetheless, this came at the cost of causing a negative impact in time-to-solution for some of our experiment configurations, with a slowdown of up to 2.5. However, this impact might be acceptable considering that we were comparing it with two architecture-specific, optimized code variants of the simulation that required extensive code refactoring to run in different architectures, and also considering the slowdown of up to 6.72 obtained with the CPU SoA OpenMP variant. Additionally, we believe that further optimizations, in particular concerning SIMD vectorization for the CPU would enhance the performance of the Kokkos variant of our code.

%Consistent with findings from previous research, as outlined in Section 2, the optimal combination of programming model and architecture for maximizing performance varies depending on the specific application. In our case, we observed better efficiencies with OpenMP on certain architectures and with Kokkos on others across our two simulations.

In Section 2, we discussed the established notion that the best programming model-architecture combination is application-specific. This is evident in our findings, where we observed better efficiencies with OpenMP on some architectures and with Kokkos on others across the two simulations we studied.

%The major limitation encountered in the codes developed for our previous paper where SoA were employed in BS-SOLCTRA for every architecture, was effectively addressed by leveraging Kokkos' polymorphic data structures. These structures ensured that regardless of the underlying architecture, the data layout remained suitable for optimal data access patterns.

The initial BS-SOLCTRA GPU version relied on one-dimensional arrays for all architectures, which limited its portability. This limitation is effectively addressed by Kokkos' polymorphic data structures, which automatically and transparently adapt the data layout to match the optimal memory access patterns for each architecture. That turned out to be a major portability advantage of the Kokkos approach over OpenMP, as it allows us maintaining a single source BS-SOLCTRA which performs well on both, CPU and GPU platforms, with an acceptable performance penalty. In a different scenario which does not require different data layouts, OpenMP is a very competitive alternative.      
%\MJR{this was not a limitation of the paper, but of our codes}

%\MJR{rephrase this, this is much too convoluted wording}
%If the trade-off of having a chance of sacrificing performance for some processing units is deemed acceptable, achieving performance %portability becomes feasible and satisfactory in plasma physics simulations with Kokkos. Alternatively, if the optimization of data %structure laying out is not considered an issue, OpenMP also offers an efficient solution to parallelize plasma physics simulation codes and %achieve performance portability across different architectures.

%and the information about the architecture optimization used by Kokkos to leverage the specific processing unit available on the target computer is accessible

% discussion
%\vspace{-1.em}
%\section{Summary and Conclusions}
\vspace{-1.em}
\section{Final Remarks}
\vspace{-.5em}
%\MJR{write a brief summary and state the main conclusions here. This section must be readable almost standalone!}
%\MJR{i.e. start with: In this paper we have presented \dots}
%\vspace{-.5em}
In this paper, we presented performance portability measurements for a field-line tracer and a particle-in-cell code, both relevant for plasma physics simulations. We used OpenMP and Kokkos variants of those two codes on CPUs from Intel, and GPUs from NVIDIA and AMD. Our Kokkos implementations achieved overall better performance portability (in terms of the formal definition in Section 2) than OpenMP, never falling below \SI{96}{\percent}, as opposed to OpenMP, which obtained a performance portability of \SI{37}{\percent} in its worst case. 
%We note that our Kokkos variants required insignificant modifications (setting the architecture, memory and execution spaces to be used when compiling the program) to execute efficiently on different architectures.
Neither the Kokkos nor the OpenMP variants suffered from significant performance drops when transitioning between processing units (GPUs or CPUs) of the same kind, \textit{e.g.} from a MI210 GPU to A100 GPU. Nevertheless, Kokkos did seem to make better use of the resources of the newer MI210, MI300A and H100 GPUs, as it achieved similar performance compared to the baseline OpenMP variant of the code with the MI210 and even outperformed the OpenMP variant on the MI300A and H100 GPUs. This contrasts with the results obtained with the older V100 and A100 GPUs on which OpenMP outperformed Kokkos. 

%\textbf{MJR: the following (1) qualifies the scope and states limitations of our study, and (2) presents some (biased) opinion and
%general advice. Feel free to disagree with (2)}

Overall, our study adds to the experience mounting in the HPC community that Kokkos and OpenMP (among others) are mature and powerful solutions for achieving performance portability of non-trivial scientific application codes (written in C++) across a broadening range of relevant HPC platforms, and - notably - they are also standing the test of time over multiple hardware generations. 
However, such studies like this by nature cannot fully address the important question \emph{``which one to choose?"}, 
developers of large-scale scientific application codes are faced with.
While the \emph{application-specific} requirements on the portability framework concerning the expressiveness, 
and flexibility to implement the relevant parallel algorithms, ease-of use, and maturity can 
be usually assessed \textit{a priori}, the \emph{strategic} question about continuing support for future HPC
platforms, and thus sustainability, remains somewhat open. Based on our experiences we are inclined to 
slightly favour open frameworks like Kokkos, as they can ultimately rely on the platform-native programming model 
for their backends, which in turn can be expected to receive earliest and highest level of support by 
the vendors.
In addition, for large code bases, the actual ``application efficiency" cannot be  judged \textit{a posteriori}, as there is no option to implement multiple variants (including platform-native programming models) and to compare relative 
performances as done here and elsewhere. Instead, comprehensive performance modelling is required to relate the 
achieved performance with the capabilities of the hardware platform which we consider an important focus for future 
performance-portability studies.

%\subsection{Key Findings}

%, considering that little to no changes in the source code were required when porting the simulations between architectures.

\vspace{0.25cm}

\noindent\textbf{Acknowledgments}. This research was partially funded by the Max Planck Society (MPG) - Costa Rica National Council of Rectors (CONARE) joint research projects framework, and was supported by machine allocations at the Costa Rica National High Technology Center (Kabré), and at the Max Planck Computing and Data Facility (Raven). We are grateful to AMD and Nvidia for providing access and consulting for their benchmark systems.

% bibliography
\bibliographystyle{splncs04}
\vspace{-1.em}
\bibliography{bib}

\end{document}